# Methylglyoxal induces cardiac dysfunction through mechanisms involving altered intracellular calcium handling in the rat heart

Hélène PEYRET [a], Céline KONECKI [a,b], Christine TERRYN [c], Florine DUBUISSON [a], Hervé MILLART [1], Catherine FELIU [a,b] and Zoubir DJERADA [a,b,*]

[a] Université de Reims Champagne Ardenne, HERVI EA 3801, Reims, 51100, France; helene.peyret@univ-reims.fr (H.P.); celine.konecki@univ-reims.fr (C.K.); florine.dubuisson@univ-reims.fr (F.D.); herve.millart@univ-reims.fr (H.M.); catherine.feliu@univ-reims.fr (C.F.); zoubir.djerada@univ-reims.fr (Z.D.);

[b] Centre Hospitalier Universitaire de Reims, Service Pharmacologie-Toxicologie, Pôle de Biologie Territoriale, Reims, 51100, France;

[c] Université de Reims Champagne Ardenne, PICT, Reims, 51100, France; christine.terryn@univ-reims.fr (C.T.);

[d] Centre Hospitalier Universitaire de Reims, Laboratoire d'Hématologie, Pôle de Biologie Territoriale, Reims, 51100, France;

* Corresponding author : Zoubir DJERADA, Université de Reims Champagne Ardenne, HERVI EA 3801, 51 rue Cognacq Jay, Reims, 51100, France. E-mail address: zoubir.djerada@univ-reims.fr (Z. DJERADA)

## ABSTRACT

Methylglyoxal (MGO) is an endogenous, highly reactive dicarbonyl metabolite generated under hyperglycaemic conditions. MGO plays a role in developing pathophysiological conditions, including diabetic cardiomyopathy. However, the mechanisms involved and the molecular targets of MGO in the heart have not been elucidated. In this work, we studied the exposure-related effects of MGO on cardiac function in an isolated perfused rat heart *ex vivo* model. The effect of MGO on calcium homeostasis in cardiomyocytes was studied *in vitro* by the fluorescence indicator of intracellular calcium Fluo-4. We demonstrated that MGO induced cardiac dysfunction, both in contractility and diastolic function. In rat heart, the effects of MGO treatment were significantly limited by aminoguanidine, a scavenger of MGO, ruthenium red, a general cation channel blocker, and verapamil, an L-type voltage-dependent calcium

Abbreviations: AUC: area under the curve; CamKII: $Ca^{2+}$/calmodulin-dependent protein kinase; DCM: diabetic cardiomyopathy; DP: developed pressure; +dp/dt: positive differentials of left ventricular force development; –dp/dt: negative differentials of left ventricular force development; HR: heart rate; LVEDP: left ventricular end-diastolic pressure; MGO: methylglyoxal; PBS: phosphate-buffered saline; RPP: rate-pressure product; RyR2: ryanodine receptor 2; SERCA2a: sarcoplasmic/endoplasmic reticulum $Ca^{2+}$-ATPase; STZ: streptozotocin; TRP: transient receptor potential; TRPA1: transient receptor potential ankyrin 1.

channel blocker, demonstrating that this dysfunction involved alteration of calcium regulation. MGO induced a significant concentration-dependent increase of intracellular calcium in neonatal rat cardiomyocytes, which was limited by aminoguanidine and verapamil. These results suggest that the functionality of various calcium channels is altered by MGO, particularly the L-type calcium channel, thus explaining its cardiac toxicity. Therefore, MGO could participate in the development of diabetic cardiomyopathy through its impact on calcium homeostasis in cardiac cells.

## KEYWORDS

## 1  INTRODUCTION

Diabetes mellitus is associated with an increased risk of cardiovascular disease [1], leading to a higher mortality risk [2]. Excessive and/or fluctuating blood glucose levels during diabetes constitute a significant factor contributing to the development of cardiovascular complications [3]. The Framingham study was the first to demonstrate that diabetic patients have a higher incidence of heart failure than a non-diabetic population [4]. However, concomitant hypertension and coronary artery disease do not explain the development of heart failure during diabetes. Indeed, diabetic patients are commonly affected by a specific type of cardiomyopathy, called diabetic cardiomyopathy (DCM). DCM is a progressive disease, beginning early after the onset of diabetes, mainly characterised by a diastolic dysfunction that occurs in the absence of coronary atherosclerosis and hypertension in diabetic patients [5]. Multiple structural myocardial changes appear in DCM, including cardiomyocyte hypertrophy and apoptosis, elevated interstitial fibrosis, and impaired coronary microvascular perfusion [6]. Unfortunately, no specific drugs are currently available for DCM treatment [7].

Methylglyoxal (MGO) is a highly reactive dicarbonyl metabolite produced during glucose metabolism. MGO is mainly produced from the fragmentation of the intermediates of glycolysis, glyceraldehyde-3-phosphate and dihydroxyacetone phosphate [8]. It may also be produced during the metabolism of lipids and proteins [9]. Some studies revealed that plasma MGO levels are significantly higher in diabetic patients due to increased dietary sugar [10–12]. Under physiological conditions, cells are protected against the toxicity of MGO due to different mechanisms, in particular, the glyoxalase system, a major detoxification pathway that converts MGO to D-lactate [13]. However, in diabetes, both hyperglycaemia and oxidative stress are associated with glutathione depletion, leading to an impairment of detoxification of MGO and worsening the diabetic state in amplification loops [14].

Because MGO is one of the most potent endogenous glycating agents, its intracellular accumulation appears highly deleterious. Some studies have demonstrated that MGO activates oxidative stress and apoptotic pathways [14–17]. Due to its high reactivity, MGO irreversibly reacts with susceptible basic amino acids on proteins to form reactive carbonyl species adducts called advanced glycation proteins (AGEs) [9]. It appears that MGO, the most abundant precursors of AGEs formation, can promote diabetic complications even in the presence of good glycaemic control [18].

Calcium is a fundamental intracellular signalling mediator in cardiac function. The alteration of calcium homeostasis is increasingly described as being involved in the cardiovascular complications of diabetes. Indeed, altered $Ca^{2+}$ current prolonged action potential duration in diabetic rat ventricular muscle [19]. Futhermore, it has been demonstrated that the $Ca^{2+}$-uptake by the sarcoplasmic reticulum was deficient in diabetic animals [20]. Cardiomyocytes isolated from rats with streptozotocin (STZ)-induced diabetes showed dyssynchronous $Ca^{2+}$ release due to a leakage of cardiac type 2 ryanodine receptors (RyR2), leading to a reduced sarcoplasmic reticulum $Ca^{2+}$ load [21]. Shao et al. identified carbonylation as a novel mechanism contributing to RyR2 dysregulation during diabetes, modifying the responsiveness of the channel to cytoplasmic $Ca^{2+}$ [22]. Through its impact on calcium homeostasis, we suggested that the accumulation of MGO during hyperglycaemia may contribute to the development of DCM. Indeed, MGO reduced sarcoplasmic/endoplasmic reticulum $Ca^{2+}$-ATPase 2a (SERCA2a) activity in STZ-induced murine model of type 1 diabetes [23]. It was also associated with diastolic dysfunction, playing a role in developing DCM [23]. Several studies revealed the involvement of transient receptor potential (TRP) channels in cardiac hypertrophy and cardiovascular remodelling [24–26]. In cardiac fibroblasts, MGO increased intracellular $Ca^{2+}$ concentration-dependent and promoted cell cycle progression from G0/G1 to S/G2/M by activating the transient receptor potential ankyrin 1 (TRPA1) channel [27]. In addition, several studies have already shown that MGO caused the entry of $Ca^{2+}$ via TRPA1 in other cell types, such as neuronal cells [28] and pancreatic beta cells [29], contributing to the development of diabetic degenerative complications and worsening diabetes. However, the underlying ionic mechanisms by which this occurs and the molecular targets of MGO remain unclear. Moreover, the impact of MGO on cardiac function has not been studied.

The present study was designed to elucidate the impact of acute exposure to MGO on cardiac function. In an *ex vivo* model of beta-1 adrenergic stimulation on Langendorff-perfused rat hearts, we studied the effects of acute exposure to MGO. Furthermore, the impact of MGO on intracellular calcium concentration was studied in isolated rat cardiomyocytes by fluorescence microscopy. The involvement of calcium channels was investigated using calcium channel inhibitors to highlight the underlying ionic mechanisms of DCM.

## 2 MATERIALS AND METHODS

### 1.1 Compounds and chemical reagents

Isoproterenol, MGO, aminoguanidine hydrochloride, ruthenium red, verapamil hydrochloride and 2,3,5-triphenyltetrazolium chloride were purchased from Sigma-Aldrich (Saint Quentin Fallavier, France).

### 1.2 Isolated heart preparation

Experiments were performed in accordance with laws and regulations controlling experiments and were approved by the local ethics committee of Reims Champagne-Ardenne n°56 (CEEA-RCA-56). All procedures complied with recommendations for animal research in France and the European Convention for the Protection of Vertebrate Animals used in Experimental and Other Scientific Purposes. All animals come from breeding in the university animal facility URCAnim. Two-month-old male Sprague-Dawley rat hearts were prepared according to the non-working Langendorff model using a retrograde perfusion system at constant pressure as previously validated and described [30]. Isolated hearts were perfused with an oxygenated Krebs-Henseleit solution at 37 °C at a constant pressure of 67 mmHg. The millimolar concentrations of constituents of the Krebs-Henseleit solution were: NaCl 118, KCl 4.7, $CaCl_2$ 1.25, $MgSO_4$ 1.2, $KH_2PO_4$ 1.2, glucose 11, $NaHCO_3$ 22.6 and ethylenediaminetetraacetic acid 0.027. These products were purchased from Sigma-Aldrich (Saint Quentin Fallavier, France).

### 1.3 Experiments

**Dose–response effect model:** After a 20-minute stabilisation period, isolated hearts were perfused with continuous treatment (t20-160 min) with MGO (Fig. 1A). Rat hearts were randomly assigned to six groups as follows: Krebs-Henseleit solution (control); 1 µM MGO; 0.1 mM MGO; 1 mM MGO; 2.5 mM MGO; or 5 mM MGO (n=4-6 rats per each group). The selected concentrations align with those reported in previous literature in other models [27,31,32]. Considering that this study is the first to expose isolated perfused rat hearts to MGO, a deliberate decision was made to test concentrations identified in earlier studies. This method was employed to characterize the dose–response relationship with cardiac function. The initial experimentation phase enabled the identification of the most effective MGO concentration for.subsequent experiments.

**Stress test model:** After a 20-minute stabilisation period, perfused hearts were submitted to stress tests with two isoproterenol stimulations at 1 µM (t30-35 min) and 10 µM (t65-70 min) for 5 minutes (Fig. 1B). Stabilisation periods were observed between each isoproterenol stimulation so that the heart could regain its baseline level of cardiac load before the following stimulation. Rat hearts were randomly assigned to six groups to receive continuous treatment (t20-115 min) as follows: Krebs-Henseleit solution (control); 1 mM MGO; 2.5 mM MGO; 1 mM MGO and 1 mM aminoguanidine; 1 mM MGO and 2 µM ruthenium red; 1 mM MGO and 0.1 µM verapamil (n=6-11 rats per each group). Treatments were prepared separately at 100X in Krebs-Henseleit solution, and perfusion flow was fixed to 1% of the mean coronary flow. The choice of MGO, aminoguanidine, ruthenium red and verapamil concentrations used in our study was based on the literature [27,33,34]. Inhibitor concentrations were adapted to our Langendorff model in order to determine the optimal concentrations of these antagonists, i.e. having the best efficacy, the greatest selectivity and the least toxicity for the heart.

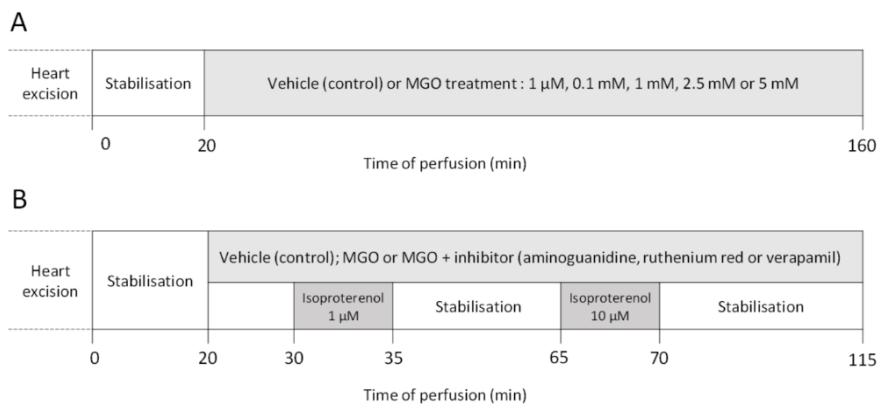

**Fig. 1.** Experimental protocol for isolated hearts perfused *ex vivo*. (A) Dose–response effect protocol. After a 20-min stabilisation period, isolated perfused hearts from 2-month-old Sprague-Dawley rats were treated with different concentrations of MGO for 140 min. (B) Stress test protocol. After a 20-min stabilisation period, isolated perfused hearts were treated with MGO alone or in combination with different antagonists such as aminoguanidine, ruthenium red or verapamil for 95 min. Alongside this, two stress tests with isoproterenol were performed, each followed by a period of stabilisation.

## 1.4 Measurements of haemodynamic parameters and necrosis

Contractile parameters: The contractile parameters were measured during the whole perfusion period. The difference between left ventricular systolic pressure (mmHg) and the left ventricular end-diastolic pressure, an index of contracture (LVEDP, mmHg), represented the left ventricular developed pressure (DP, mmHg). The heart rate (HR, beats·min$^{-1}$), the positive differentials of left ventricular force development (+dP/dt, mmHg·min$^{-1}$) and the negative differentials of left ventricular force development (–dP/dt, mmHg·min$^{-1}$) were measured simultaneously. The measured values are recorded every 10 seconds. The rate-pressure product (RPP, mmHg·beat·min$^{-1}$) was calculated by multiplying DP and HR.

Mean coronary flow: During the perfusion, the mean coronary flow was measured using the perfusate draining out for 1 min and normalised to the wet heart weight (mL·min$^{-1}$·g$^{-1}$).

Necrosis volume determination: After 2,3,5-triphenyltetrazolium chloride staining, tissular necrosis volume was determined at the end of the perfusion by quantitative image analysis as previously described [30].

## 1.5 Cell culture

According to the manufacturer's instructions, neonatal cardiomyocytes were isolated from 1–4-day-old Sprague Dawley rats using the Neonatal Cardiomyocyte Isolation System (Worthington Biochemical Corporation, Reading, UK). Cardiomyocytes were cultured in Leibovitz's L-15 medium (Worthington, Biochemical Corporation, Reading, UK) supplemented with 5% (vol/vol) foetal calf serum (Dominique Dutscher, Bernolsheim, France) and 1% (vol/vol) penicillin and streptomycin solution (Life Technologies; Invitrogen, Cergy-Pontoise, France).

## 1.6 Measurement of intracellular calcium levels

Neonatal rat cardiomyocytes in culture were used to quantify cytosolic free calcium levels using Fluo-4 Direct™ Calcium Assay Kit (Molecular Probes, Life Technologies, Eugene, USA) by automatised fluorescent microscopy Axio observer Z1 (CarlZeiss, Marly-le-Roi, France) with coolSNAP HQ2 camera (Roper Scientific, Lisses, France). The acquisition was performed by the software MetaMorph v7.8 (Roper Scientific, Lisses, France). Adherent cardiomyocytes cultured in µ-Slide 8-well ibiTreat (Ibidi, Clinisciences, Nanterre, France) were loaded with Fluo-4 for 45 min at 37°C under 5% $CO_2$. Kinetics of fluorescence intensity was measured with a frequency of 1 image per second at 37°C and 5% $CO_2$, using an excitation wavelength of 475 nm and an emission of 530 nm. In the first experiment, during 15 min acquisition, MGO (100, 200, or 500 µM) or phosphate-buffered saline (PBS; Control) was added after 1 min of stabilisation.

Secondly, during 17 min acquisition, ruthenium red (10 µM) or verapamil (1 µM), two calcium channels inhibitors, were added after 1 min of stabilisation. Then MGO (500 µM) or PBS (Control) was added at t3 min in the observed well. The fluorescence intensity analysis was performed with ImageJ version 1.49 (National Institutes of Health, USA). Relative fluorescence intensity (%) measurement was normalised to the average baseline value before pharmacological treatment.

### 1.7 Statistical analysis

Statistical analyses were performed using GraphPad Prism version 6.0 for Windows (GraphPad Software, San Diego, California USA). All values of experiments are expressed as mean ± SEM (Standard Error of Mean). We previously verified that the data were distributed according to a normal distribution with Kolmogorov-Smirnov and Shapiro-Wilk test. Subsequently, we performed statistical analyses a single-factor repeated ANOVA test for kinetic data and a single-factor ANOVA test for other data, followed by the Tukey post hoc test to highlight significant differences ($p < 0.05$) between groups. When data were not distributed according to the normal distribution, statistical analyses were performed using the Friedman test for kinetics data and the Mann-Whitney test for other data to highlight significant differences ($p < 0.05$) between groups.

## 2 RESULTS

### 2.1 MGO induced concentration-dependent contractile parameters impairment, coronary dysfunction, and tissue necrosis in isolated perfused rat heart

In the initial exploration phase, after 20 min of stabilisation, perfused isolated hearts were continuously exposed to increasing concentrations of MGO (1 µM, 0.1 mM, 1 mM, 2.5 mM, or 5 mM) for 140 min (Fig. 1A). At low concentrations of 1 µM and 0.1 mM, MGO had no significant effect on relative DP, HR, and RPP, compared to the control (Figs. 2A–C). However, there was a slight but significant increase in +d$p$/dt and –d$p$/dt with 0.1 mM MGO compared to the control group (at t160 min, 98.0 % ± 3.2 % and 87.2 % ± 2.5 %, respectively, for the 0.1 mM MGO group, vs. 85.5 % ± 7.9 % and 76.3 ± 6.3 % for the control group, $p < 0.05$) (Supplementary data, Figs. S1A and B). In the 1 mM MGO group, during the entire protocol, relative DP was significantly increased up to 142.2 % ± 9.9 % at t160 min compared to the control group, which remained stable and then gradually decreased to 81.3 % ± 4.9 % ($p = 0.002$) (Fig. 2A). Except for HR, most of the contractile parameters changed in the same way: +d$p$/dt, –d$p$/dt and RPP were significantly increased in the 1 mM MGO group compared to the control group ($p < 0.05$). On the contrary, above 1 mM concentration, MGO induced a decrease in cardiac parameters. Indeed, in the 2.5 mM MGO group, HR and RPP decreased gradually and significantly compared to the control group ($p < 0.05$). The 5 mM MGO group showed a drastic and significant fall in all contractile parameters, DP, HR, +d$p$/dt, –d$p$/dt and RPP, leading to a condition similar to cardiac arrest ($p < 0.01$).

MGO showed a concentration-dependent effect on LVEDP of perfused hearts (Fig. 2D). The higher the MGO concentration, the greater the increase in LVEDP compared to the control group, except for the lowest concentration of 1 µM which did not show a significant increase. Indeed, at the end of the time course, LVEDP was drastically increased by MGO at 2.5 mM and 5 mM (27.5 mmHg ± 5.1 mmHg and 35.9 mmHg ± 4.8 mmHg, respectively, vs. 12.5 mmHg ± 2.8 mmHg for the control group, $p < 0.001$). In perfused hearts treated with lower concentrations of 0.1 mM and 1 mM, the increase in LVEDP was smaller but still significantly higher than the control ($p < 0.05$).

MGO at the concentration from 1 µM to 2.5 mM had no significant effect on the coronary flow (Fig. S1C). In contrast, MGO at 5 mM showed a significant decrease compared to the control group: at t160 min 44.3 % ± 1.4 % for the 5 mM MGO group vs. 105.7 % ± 1.4 % for the control group ($p = 0.049$).

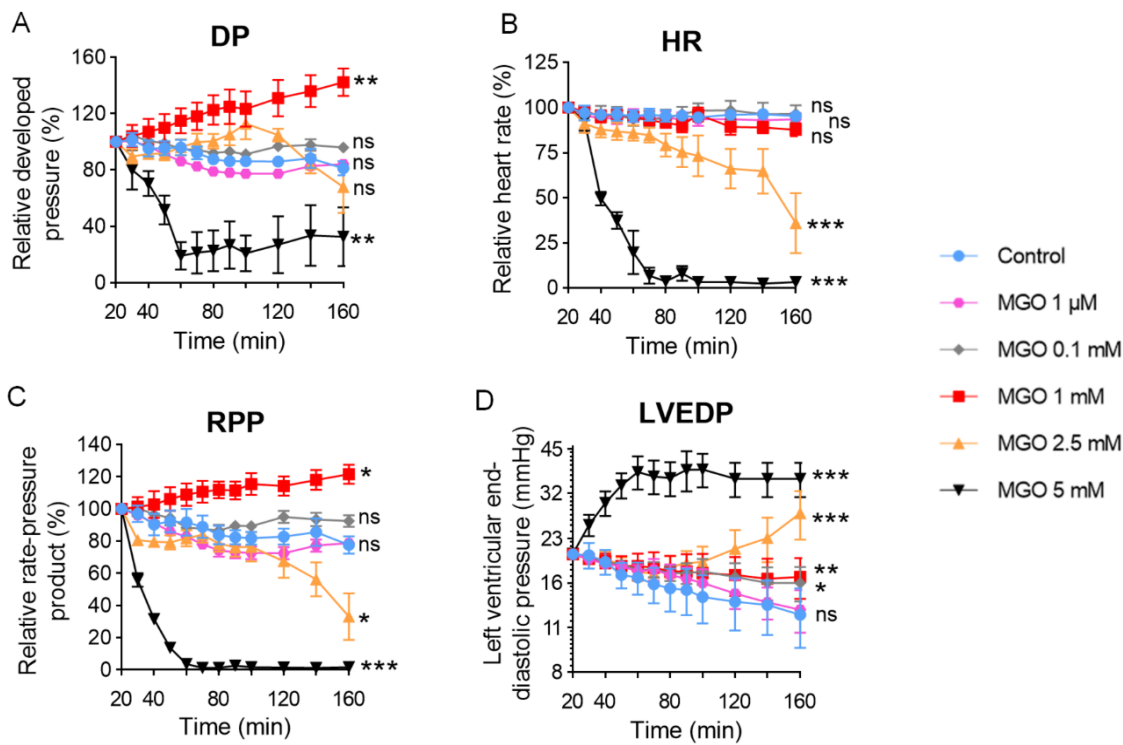

**Fig. 2.** Dose-effect of MGO on cardiac function. After 20 min of stabilisation, isolated perfused hearts were exposed to MGO (1 µM, 0.1 mM, 1 mM, 2.5 mM or 5 mM) or vehicle control for 140 min. Cardiac function was determined by measurement of developed pressure (DP) (A), heart rate (HR) (B), rate-pressure product (RPP) (C), left ventricular end-diastolic pressure (LVEDP) (D) in percentage relative to baseline. Means ± SEM of 4-6 rats per groups. * $p < 0.05$; ** $p < 0.01$; *** $p < 0.001$: significantly difference from control group by Friedman comparison test.

Histological staining of the hearts with TTC showed that the higher the MGO concentration, the greater the volume of necrosis (Fig. 3). Indeed, the volume of necrosis was significantly increased in the 2.5 mM and 5 mM MGO groups compared to the untreated control hearts (13.3 % ± 3.8 % p = 0.003, 21.3 % ± 4.6 % p = 0.002, respectively, vs. 3.4 % ± 0.5 % for the control group) while this increase was no significant in the 1 mM MGO group (6.0 % ± 1.5 % p = 0.112). The lower concentrations 1 µM and 0.1 mM showed no effect on the tissue necrosis.

The increase of DP in the hearts treated by MGO at 1 mM suggested that MGO induced calcium overload without severe toxicity on cardiac function in conjunction with moderate markers of suffering. Above 1 mM, MGO appeared to induce severe toxicity leading to an overall decrease in functional parameters and a significant increase in tissue necrosis volume. For this reason, the highest concentration 5 mM was excluded for the stress test model of isolated perfused hearts.

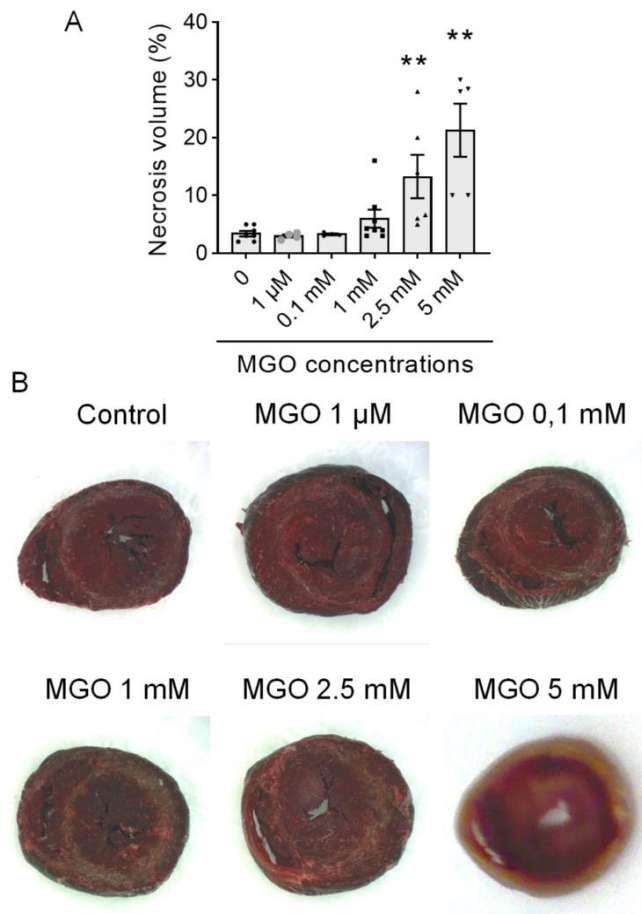

**Fig. 3.** MGO increased necrosis volume in isolated perfused rat hearts. At the end of 140 min of MGO treatment, hearts were stained with TTC, fixed and sliced to determine necrosis volume. (A) Quantification of necrotic tissue (white area) of TTC-stained hearts and (B) representative photographs of heart sections. Means ± SEM of 4-6 rats per groups; ** $p < 0.01$: significantly difference from control by Mann-Whitney comparison test.

## 2.2   Cardiac dysfunction induced by MGO involved altered calcium regulation

In perfused hearts treated with MGO at 1 mM or 2.5 mM, after each isoproterenol stimulation, the DP took longer to return to baseline than in control (Fig. 4). For 1 µM and 10 µM isoproterenol stimulations, the control returned to baseline after 10 min and 24 min, respectively. In the 1 mM MGO group, DP decay occurred after 25 min and 40 min and in the 2.5 mM MGO group, it decreased after 30 min and 45 min. Thus, to highlight this decay time, the area under the curve (AUC) was calculated by summing the measured values over time during each post-stimulation stabilisation period from 35 to 65 min for 1 µM isoproterenol stimulation, and from 70 to 115 min for 10 µM isoproterenol stimulation. The AUC data of 10 µM isoproterenol-stimulation were reported in Figure 5 and Supplementary data (Fig. S2). The AUC data of 1 µM isoproterenol-stimulation are available in Supplementary data (Fig. S3) and showed similar results.

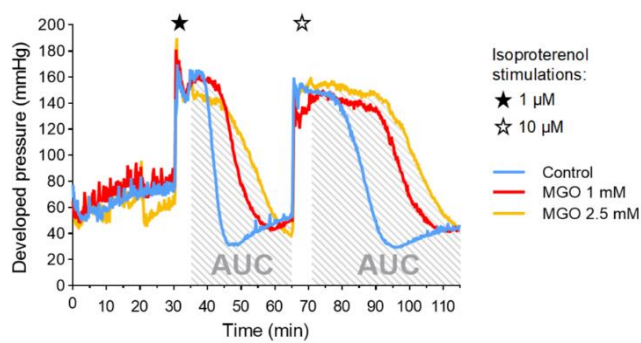

**Fig. 4.** Effect of MGO on the developed pressure (DP) of isolated perfused rat heart stimulated with isoproterenol 1 µM then isoproterenol 10 µM. The area under the curve (AUC) was calculated during all stabilisation periods after each isoproterenol stimulation.

The 1 mM and 2.5 mM MGO groups showed a significant increase of the AUC of DP ($2.6 \times 10^4 \pm 0.8 \times 10^3$ and $3.1 \times 10^4 \pm 1.7 \times 10^3$ mmHg·min, respectively) compared to control ($1.9 \times 10^4 \pm 1.0 \times 10^3$ mmHg·min, $p < 0.001$; Fig. 5A). Differently, MGO at 1 mM had no significant effect on HR, whereas MGO at 2.5 mM significantly decreased the AUC of HR compared to control ($7.1 \times 10^5 \pm 2.3 \times 10^3$ beats·min$^{-1}$·min for the 2.5 mM MGO group vs. $8.0 \times 10^5 \pm 1.3 \times 10^3$ beats·min$^{-1}$·min for the control group, $p = 0.001$; Fig. 5D). In addition, the AUC of RPP in the 1 mM MGO group was increased compared to the control ($p < 0.001$) but not in the 2.5 mM MGO group ($p = 0.088$) (Fig. 5B). As RPP is the product of DP and HR, the opposing effects of MGO at 2.5 mM on the AUC of DP and HR compensate each other in the AUC of RPP. Moreover, in the 1 mM and 2.5 mM MGO groups, the AUC of +d$p$/dt and –d$p$/dt was increased compared to the control group ($p < 0.001$). Furthermore, MGO at 1 mM and 2.5 mM maintained the mean of LVEDP to high values compared to control during the whole perfusion ($21.5 \pm 0.7$ mmHg and $20.4 \pm 0.2$ mmHg respectively, vs. $17.9 \pm 0.4$ mmHg in the control group, $p < 0.001$) (Fig. 5C).

The increase in the AUC of contractile parameters induced by MGO at 1 mM was effectively limited by 1 mM aminoguanidine, an MGO scavenger, 2 µM ruthenium red, a general antagonist of calcium channels, and 0.1 µM verapamil, a $Ca^{2+}$ L-type channel inhibitor (Figs. 5A and B, S2A and B). Indeed, the AUC of DP was significantly reduced in the MGO + aminoguadinine, MGO + ruthenium red, and MGO + verapamil groups ($2.1 \times 10^4 \pm 1.8 \times 10^3$, $2.0 \times 10^4 \pm 2.2 \times 10^4$ and $1.7 \times 10^4 \pm 1.2 \times 10^3$ mmHg·min, respectively) compared to the 1 mM MGO group ($2.6 \times 10^4 \pm 0.8 \times 10^3$ mmHg·min, $p < 0.01$), and was not significantly different from the control group ($1.9 \times 10^4 \pm 1.0 \times 10^3$ mmHg·min, $p > 0.05$; Fig. 5A). As above, aminoguanidine, ruthenium red and verapamil significantly limited the increase in the AUC of +d$p$/dt, −d$p$/dt and RPP induced by MGO ($p < 0.01$; Figs. 5B and S2). These inhibitors also significantly reduced the MGO-induced increase of LVEDP ($17.5 \pm 1.2$ mmHg, $15.3 \pm 1.6$ mmHg, $18.5 \pm 0.7$ mmHg, respectively, compared to $21.5 \pm 0.7$ mmHg in the MGO 1 mM group, $p < 0.05$) (Fig. 5C).

The coronary flow measured at the end of the procedure was also increased by MGO at 1 mM and 2.5 mM ($17.1 \pm 0.5$ mL·min$^{-1}$ and $18.0 \pm 0.6$ mL·min$^{-1}$, respectively) compared to the control group ($11.9 \pm 0.5$ mL·min$^{-1}$; $p < 0.001$) (Fig. S3F). This increase of the coronary flow, explained by the higher contractility in MGO-treated hearts, was effectively limited by aminoguanidine, ruthenium red and verapamil ($p < 0.05$).

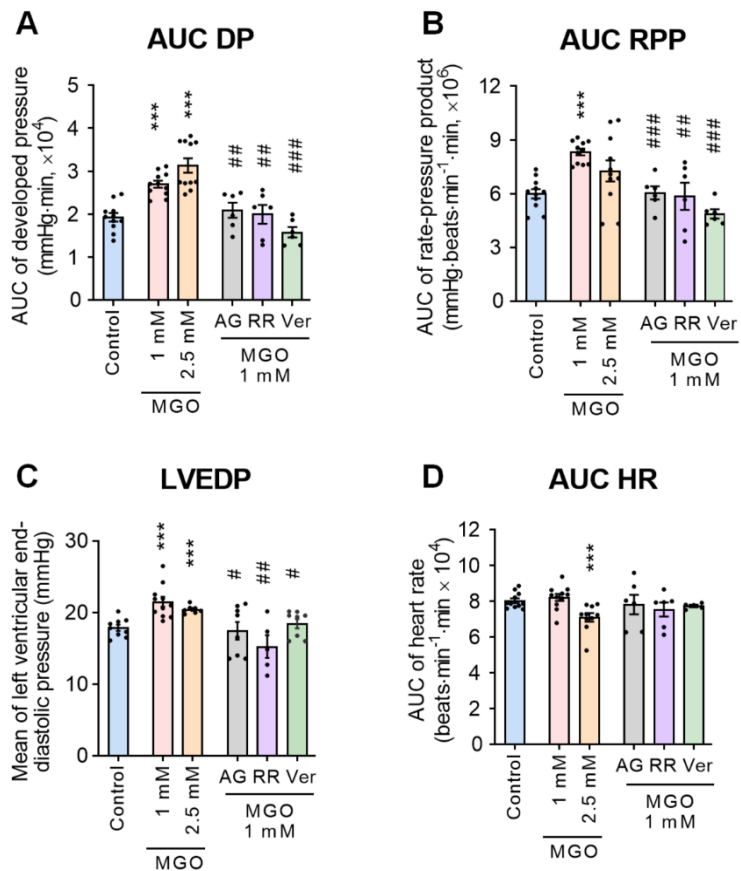

**Fig. 5.** Effect of MGO on cardiac haemodynamic response to stress test after 10 µM isoproterenol stimulation in isolated perfused hearts. Data shows the area under curve of each functional parameter, developed pressure (DP) (A), rate-pressure product (RPP) (B), heart rate (HR) (D), and mean of left ventricular end-diastolic pressure (LVEDP) (C) during the stabilisation period after the second stress test with 10 µM isoproterenol. After 20 min of stabilisation, hearts received either perfusate alone vehicle control; 1 mM MGO alone; 2.5 mM MGO alone; 1 mM MGO + 1 mM aminoguanidine (AG); 1 mM MGO + 2 µM ruthenium red (RR); or 1 mM MGO + 0.1 µM verapamil (Ver) for 95 min. Means ± SEM of 6-11 rats per group. *** $p < 0.001$: significantly different from control and # $p < 0.05$; ## $p < 0.01$; ### $p < 0.001$: significantly different from MGO 1 mM group; by Mann-Whitney comparison test.

## 2.3 MGO increased intracellular $Ca^{2+}$ levels in newborn rat cardiomyocytes in a concentration- and time-dependent manner

To investigate the effect of MGO on calcium handling in cardiomyocytes, the intracellular $Ca^{2+}$ levels assessment using Fluo-4 probes was applied. MGO significantly increased intracellular $Ca^{2+}$ levels in a

concentration- and time-dependent manner (Fig. 6). In the 100 µM and 200 µM MGO groups, intracellular $Ca^{2+}$ levels increased progressively and significantly compared to the control (at T15 min, 157.1 % ± 16.4 % and 162.9 % ± 16.1 %, respectively, vs. 112.0 % ± 9.0 % in the control group, $p < 0.01$). At 500 µM, MGO induced a sharp increase in intracellular $Ca^{2+}$ levels, reaching 123.2 % ± 11.2 %, only 1 min after the addition of MGO, followed by a slight decrease to 116.1 % ± 9.4 % at T3 min. Subsequently, the 500 µM MGO group showed a strong increase in intracellular $Ca^{2+}$ levels reaching 173.2 % ± 12.7 % at the end of acquisition ($p < 0.001$).

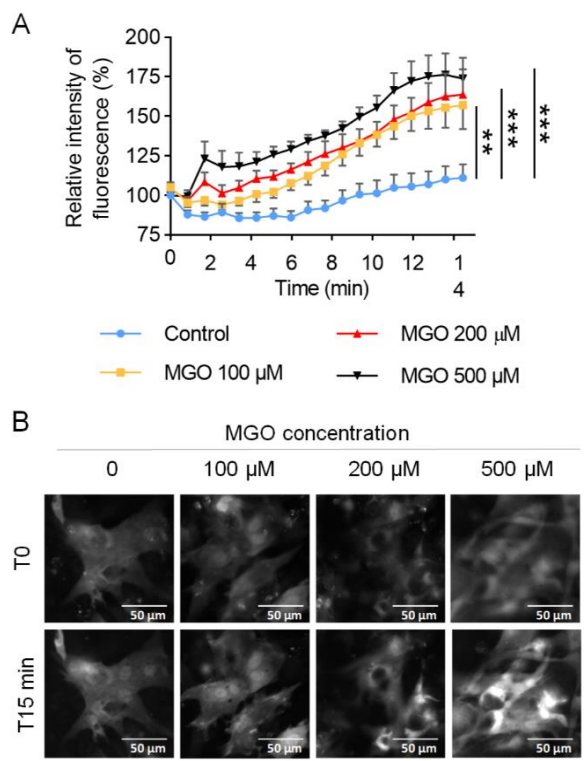

**Fig. 6.** Effect of MGO on intracellular calcium levels in newborn rat cardiomyocytes. After 45 min-incubation with Fluo-4, fluorescence intensity was measured in fluorescence microscopy (ex 475 nm/em 530 nm). (A) After 1 min of acquisition, MGO was added in medium to obtain 100, 200 or 500 µM concentration and fluorescence measurement was continued for 14 min (means of 7-9 wells ± SEM; *** $p < 0.001$, ** $p < 0.01$ vs. control). Comparisons were made using a two-way ANOVA test. (B) Representative fluorescent microscopy images of intracellular calcium levels in newborn rat cardiomyocytes before (T0) and after 14 min MGO-treatment (T15 min).

## 2.4 MGO-induced intracellular $Ca^{2+}$ increase was limited by aminoguanidine and calcium channels inhibitor

After that we investigated the effects of aminoguanidine, a methylglyoxal scavenger, ruthenium red and verapamil, two calcium channels inhibitors, on MGO-induced intracellular $Ca^{2+}$ increase (Fig. 7). Aminoguanidine and verapamil significantly limited the effects of MGO on intracellular $Ca^{2+}$ levels (at T17 min, 136.1 % ± 18.7 % and 133.3 % ± 13.0 %, respectively, vs. 186.5 ± 11.7 % in the MGO group, p < 0.05) to values that are not significantly different from the control group (138.1 % ± 17.3 %) (Fig. 7A). On the other hand, ruthenium red failed to limit MGO-induced $Ca^{2+}$ increase (at T17 min, 177.5 % ± 20.0 %, p = 0.694 vs. MGO group). However, this calcium channel antagonist showed a slight decrease, but not significant, in intracellular $Ca^{2+}$ levels throughout the kinetics (Fig. S4).

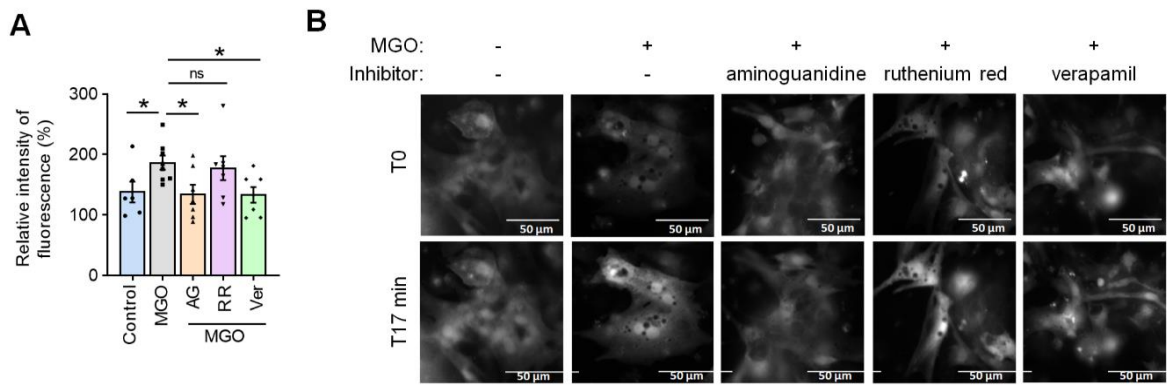

**Fig. 7.** Effect of aminoguanidine and calcium channels inhibitors on MGO-induced intracellular calcium levels increase. After 45 min-incubation with Fluo-4, fluorescence intensity was measured in fluorescence microscopy (ex 475 nm/ex 530 nm). After 1 min of acquisition, aminoguanidine (AG) at 500 µM, ruthenium red (RR) at 10 µM or verapamil (Ver) at 1 µM was added to the medium, then at the third minute, MGO at 500 µM was added, and the acquisition was continued for 14 min. (A) Comparisons were made between fluorescence intensity of the endpoint T17 min of each groups (means of 6-8 wells ± SEM; * p < 0.05) using Mann-Whitney test. (B) Representative fluorescent microscopy images of intracellular calcium levels in newborn rat cardiomyocytes before (T0) and at after 14 min MGO-treatment (T17 min).

## 3  DISCUSSION

In the present study, we found that MGO, at 1 mM, induced cardiac dysfunction with an increase in DP, RPP, +d$p$/dt and –d$p$/dt as well as in LVEDP. Calcium plays an essential role in the coupling excitation-contraction-relaxation mechanisms. This increase in cardiac contractility and the development of a contracture as shown by LVEDP could reflect the occurrence of a calcium overload and the overcoming of the mechanisms required for the extrusion of calcium to the extracellular environment. The same cardiac dysfunctions are aggravated during the isoproterenol-induced stress test. Ruthenium red and verapamil, two calcium channel inhibitors were able to limit MGO-induced cardiac dysfunction. Our experimental data at the cellular level confirm this MGO-induced calcium overload in a concentration- and time-dependent manner. These effects were limited by verapamil which demonstrates the involvement of the calcium channels and most likely the L-type calcium channel.

Diabetes significantly increases endogenous MGO, a dicarbonyl reactive compound, which may play a role in developing diabetes-related diseases, including DCM. Initially, this work consisted in studying the effect of different concentrations of MGO on cardiac function in the Langendorff model. This *ex vivo* technique allowed us to assess the direct impacts of MGO without the complications associated with *in vivo* models and by avoiding the confounding effect of the neuronal and hormonal environment of the living animal [34,35]. In this first exploratory experiment, the toxicity of MGO described a concentration-dependent effect relationship. At 1 mM, MGO appeared to increase DP, +d$p$/dt and –d$p$/dt, reflecting strength and speed of contractility. This MGO-induced increase in contractility is the first indication of calcium entry in the heart. This hypothesis is supported by previous studies showing that MGO induces intracellular calcium accumulation in cardiac fibroblasts [27] and in vascular smooth muscle cells [36]. In complete opposition, at 2.5 mM and 5 mM, MGO has significant deleterious effects on cardiac function by reducing all functional parameters. MGO at 5 mM even showed evidence of acute cytotoxicity leading to diastolic dysfunction, coronary flow loss, and extensive tissue necrosis. This reduction in all functional parameters by the higher concentrations of MGO could be caused by various mechanisms described in the literature. Indeed, it has been shown that MGO induced apoptosis and severe oxidative stress in rat cardiomyocytes [22,31] and, by reacting with a large number of protein sites, could modify cardiomyocyte myofilament reducing their calcium sensitivity and maximal calcium-activated force [37].

Given this observation, we wanted to further investigate the results of this experiment by subjecting the isolated perfused hearts to beta-1 adrenergic stimulations to study the impact of MGO on cardiac response during the stress tests. As isoproterenol stimulation is transient and reversible, each heart was stimulated twice with 1 and 10 µM isoproterenol with a stabilisation period between each stimulation to allow the heart to recover basal function. In this upcoming experiment, we will focus exclusively on the 1 mM and 2.5 mM concentrations of MGO. This choice is based on our observation of noticeable effects starting at the 1 mM concentration in the earlier ex vivo dose-response model (Fig 2). Furthermore, it was observed that the 5 mM concentration resulted in significant toxicity, affecting both functional parameters and causing tissue damage (Fig 2 and 3).

The results showed that 1 mM MGO induced a sustained positive inotropic effect of isoproterenol. This sustained positive inotropic effect was worsened by 2.5 mM MGO. This delay in the return of developed pressure to basal level suggests that MGO caused dysfunction in intra- and/or extra-cellular calcium fluxes during beta-1 adrenergic receptor signalling pathways. Moreover, MGO-treated hearts showed elevated LVEDP, a marker of ventricular contracture, suggesting a possible calcium overload leading to left ventricular diastolic dysfunction. These results are consistent with previous studies showing a link between MGO and cardiac dysfunction. Firstly, the glyoxalase I overexpression, a MGO detoxification enzyme, delayed and limited the loss of cardiac function in diabetic mouse [38]. Secondly, a cohort study with long-term type 1 diabetic patients established a link between MGO and DCM with a correlation between the circulating concentration of MGO-derived hydroimidazolone and left ventricular diastolic dysfunction [39]. Moreover, the early stages of DCM are commonly characterised by left ventricular diastolic dysfunction [40].

To determine the specificity of dysfunction induced by MGO, the first approach was to antagonise the effects of MGO by aminoguanidine, whose nucleophilic character gives it the property of reacting readily with aldehydes and ketones [41]. Many studies have demonstrated its effectiveness in MGO inhibition *in vivo* and *in vitro* [42,43]. In our study, aminoguanidine effectively inhibited the cardiac dysfunction MGO-induced in the isolated perfused rat heart model. This result confirmed the specific impact of MGO on cardiac function.

The second approach was to use two calcium channel blockers to determine whether the MGO-induced cardiac dysfunction involved impaired calcium handling in the heart. First, we used ruthenium red, a general antagonist of calcium channels, including the voltage-dependent calcium channels [44], the mitochondrial calcium uniporter [45,46], and which can also block the calcium uptake and release from the sarcoplasmic reticulum [47]. Ruthenium red was also used for its ability to antagonise TRP channels, whose involvement in MGO-induced calcium entry into cardiac fibroblasts has recently been demonstrated [27]. Furthermore, verapamil, an L-type voltage-dependent calcium channel blocker, reduces the influx of calcium ions through the cell membrane, which occurs when the cell is depolarised. Therefore, verapamil was used to determine whether the calcium fluxes generated by over-activation of L-type calcium channel were involved in MGO-induced dysfunctions. In our work, the inhibitory effect of ruthenium red and verapamil highlighted the leading role of calcium flux and the involvement of calcium channels in MGO-induced cardiac dysfunction. The effectiveness of verapamil in limiting MGO-induced cardiac dysfunction proves that this dysfunction involved especially the L-type calcium channel. Also, it is worth noting that the DP is more affected by 1 mM MGO than HR, suggesting that the toxicity of MGO mainly targets calcium regulation rather than cardiac electrical activity as subjected by no significant change in HR.

To elucidate the involvement of calcium regulation in the observed effects of the ex vivo model, a complementary in vitro model using newborn rat cardiomyocytes was proposed. MGO concentrations, which had shown potential effects in previous ex vivo experiments, were selected. Initial tests indicated noticeable effects starting at 100 µM. This range aligns with concentrations reported in the literature [27,31], leading to the use of MGO concentrations between 100 µM and 500 µM in the cardiomyocyte culture model. Furthermore, it was ensured that these concentrations did not induce cell death. Findings revealed that cell viability was not compromised after 24 hours of exposure to methylglyoxal at concentrations up to 500 µM (Figure S5). This outcome provided confidence that a brief exposure of 15 to 17 minutes to MGO at concentrations ranging from 100 to 500 µM would not result in cytotoxic effects.

In newborn rat cardiomyocyte culture, MGO induced calcium entry in a concentration-dependent manner. This results is in line with many studies which have shown that MGO increased intracellular calcium levels in other cell types as pancreatic cells [48], cardiac fibroblasts [27], renal tubular cells [49], immortalised cell line HL-1 [31], and neuronal cells [28]. In our study, the increase of intracellular calcium levels in cardiomyocytes was completely limited by verapamil, demonstrating the involvement of the L-type calcium channel. However, in this *in vitro* cell model, ruthenium red was less effective than in the previous *ex vivo* perfused heart model, only showing a non-significant downward trend in intracellular calcium levels.

Together, these results showed that MGO leads to dysregulation in intra- or extra-cellular calcium fluxes via calcium channels sensitive to ruthenium red and verapamil. The impairment of calcium-handling in cardiomyocytes could be involved in the MGO-induced cardiac dysfunction, which be demonstrated for the first time in the isolated perfused heart model.

Several animal and human models have shown that high blood glucose levels during diabetes predispose to dysfunctional calcium homeostasis [50]. This dysregulation of intracellular calcium was associated with diastolic dysfunction [51,52], prolonged depolarisation [52,53]. The accumulation of intracellular calcium contributed to the chronic activation of calcium-dependent proteins, such as $Ca^{2+}$/calmodulin-dependent protein kinase II (CaMKII), leading to apoptosis [54], fibrosis [55], and hypertrophy [56], then resulting in heart failure [57,58]. Although these studies did not directly include MGO in their research strategy, the accumulation of toxic diabetes metabolites such as MGO in these hyperglycaemic models is very plausible. Therefore, the toxicity of hyperglycaemia would be mediated at least in part by MGO, which is in line with our model and corroborates the results obtained in the present work. In these diabetes-induced cardiac calcium overload models, calcium channel blockers appeared to have protective effects. Indeed, azelnidipine, a long-acting calcium channel blocker, alleviated STZ-induced myocyte contractile dysfunction [59]. Moreover, verapamil prevented cardiomyocyte apoptosis [60], left-ventricular dysfunction and myocardial changes in diabetic mice [61]. As in these diabetic models, in our study, verapamil limited the impact of MGO on calcium homeostasis, demonstrating the key role of calcium channels as a therapeutic target in DCM.

The present study highlighted the impact of acute exposure to MGO on cardiac function, most likely involving membrane calcium channels. Some studies have shown that MGO could also impact the different components of the regulation of intracellular calcium in various cell types, such as the ryanodine receptors [22] and the SERCA2a [23,62]. In the perspective of this work, the involvement of other components of calcium regulation could be studied using selective blockers. The role of CaMKII in MGO-induced cardiac dysfunction should be explored [58]. Moreover, to get closer to the pathophysiological conditions of diabetes, it would be interesting to study the impact of MGO on cardiac function in a chronic phase. Other biguanides close to aminoguanidine could also be tested, such as metformin which is currently used an anti-diabetic as first line treatment in humans and known to be cardioprotective.

In conclusion, our study demonstrates the direct impact of MGO on systolic and diastolic cardiac functions in relation to the impairment of calcium homeostasis. In our model, the alteration of calcium homeostasis is most likely related to the dysfunction of calcium channels, in particular the L-type voltage-dependent calcium channel. Calcium overload partly explains the impact of MGO on cardiac function in diabetic patients.

## APPENDIX A. SUPPLEMENTARY DATA

Supplementary data to this article can be found in the Supplementary date file.

## ACKNOWLEDGEMENTS

The authors thank Mr Anthony Pigeon and Mrs Perrine Manot for raising et caring for the animals in the URCAnim platform, Mrs Marie Traoré for administrative support, Mrs Floriane Oszust and Mrs Florine Dubuisson for her assistance and expertise. The authors thank Prof Mulder and Dr Larabi for their advices and expertise.

## FUNDING SOURCES

This work was supported by the University of Reims Champagne-Ardenne.

# Supplementary data

**Methylglyoxal induces cardiac dysfunction through mechanisms involving altered intracellular calcium handling in the rat heart**

Hélène Peyret[a], Céline Konecki[a,b], Christine Terryn[c], Florine DUBUISSON [a], Hervé Millart[1], Catherine Feliu[a,b] and Zoubir Djerada[a,b,*]

[a] Université de Reims Champagne Ardenne, HERVI EA 3801, 51 rue Cognacq Jay, Reims, 51097, France;

[b] Centre Hospitalier Universitaire de Reims, Service Pharmacologie-Toxicologie, Pôle de Biologie Territoriale, rue du Général Koenig, Reims, 51100, France;

[c] Université de Reims Champagne Ardenne, PICT, 51 rue Cognacq Jay, Reims, 51097, France;

[d] Centre Hospitalier Universitaire de Reims, Laboratoire d'Hématologie, Pôle de Biologie Territoriale, rue du Général Koenig, Reims, 51100, France;

* Corresponding author. Université de Reims Champagne Ardenne, HERVI EA 3801, 51 rue Cognacq Jay, Reims, 51097, France. E-mail address: zoubir.djerada@univ-reims.fr.

**Summary**

- **Supplementary Table S1**
- **Supplementary Figure S1**
- **Supplementary Figure S2**
- **Supplementary Figure S3**
- **Supplementary Figure S4**
- **Supplementary Figure S5**
- 
- **Supplementary File :**

    Attached file « Time-lapse_video_Fluo-4_cardiomyocytes_20X_ionomycin.mpg »;

    Time-lapse video of cardiomyocytes loaded with Fluo-4 and treated by ionomycin at 5 µM.

    Acquisition using fluorescence microscopy (ex 475 nm/em 530 nm) at 1 image per second. Video speed has been multiplied by 20.

**Table S1.** Absolute values of haemodynamic parameters at the time t20 minutes. Developed pressure (DP), heart rate (HR), rate-pressure product (RPP), positive differentials of left ventricular force development (+d$p$/dt), negative differentials of left ventricular force development (−d$p$/dt), and coronary flow. Means of 4-6 rats per groups. * $p < 0.05$: significantly difference from control group by Mann-Whitney comparison test.

|  | Control | MGO 1 μM | MGO 0.1 mM | MGO 1 mM | MGO 2.5 mM | MGO 5 mM |
|---|---|---|---|---|---|---|
| DP (mmHg) | 72.50 | 55.00 | 64.50 | 57.17 | 79.83 | 63.20 |
| HR (BPM·min$^{-1}$) | 264.8 | 277.3 | 247.5 | 254.7 | 249.7 | 219.4 |
| RPP (mmHg·BPM·min$^{-1}$) | 19347 | 15140 | 15631 | 14877 | 18625 | 18979 |
| +d$p$/dt (mmHg·min$^{-1}$) | 2432 | 1897 | 2049 | 1942 | 2496 | 2599 |
| −d$p$/dt (mmHg·min$^{-1}$) | 1719 | 1375 | 1523 | 1386 | 1947 | 2100 |
| Coronary flow (mL·min$^{-1}$) | 14.17 | ***18.50**** | 17.00 | 14.50 | 14.33 | 14.00 |

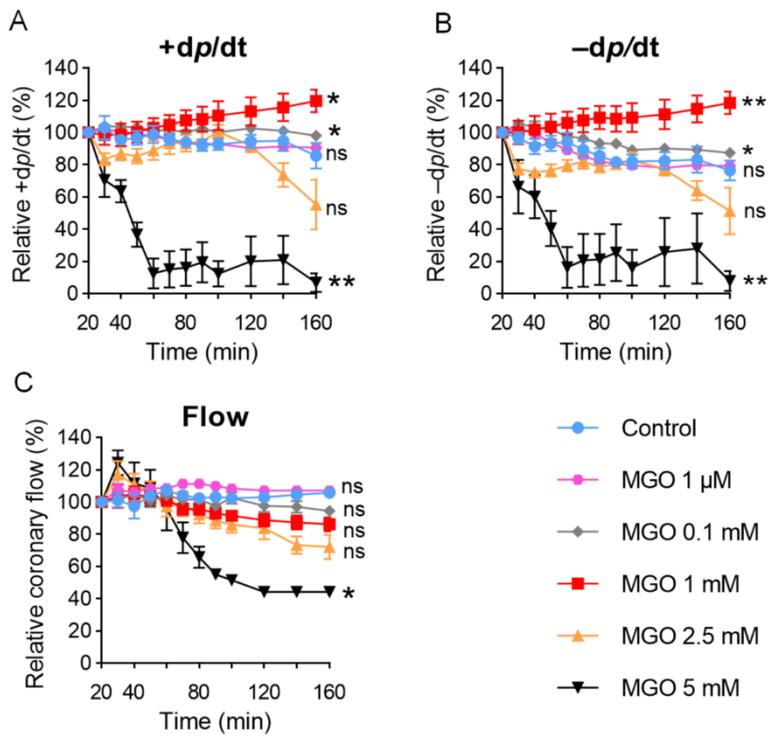

**Fig. S1.** Dose-effect of MGO on cardiac function. After 20 min of stabilisation, isolated perfused hearts were exposed to MGO (1 µM, 0.1 mM, 1 mM, 2.5 mM or 5 mM) or vehicle control for 140 min. Positive differentials of left ventricular force development (+d$p$/dt) (A), negative differentials of left ventricular force development (−d$p$/dt) (B), and coronary flow (C) are expressed in percentage relative to baseline. Means ± SEM of 4-6 rats per groups. * $p < 0.05$; ** $p < 0.01$: significantly difference from control group by Friedman comparison test.

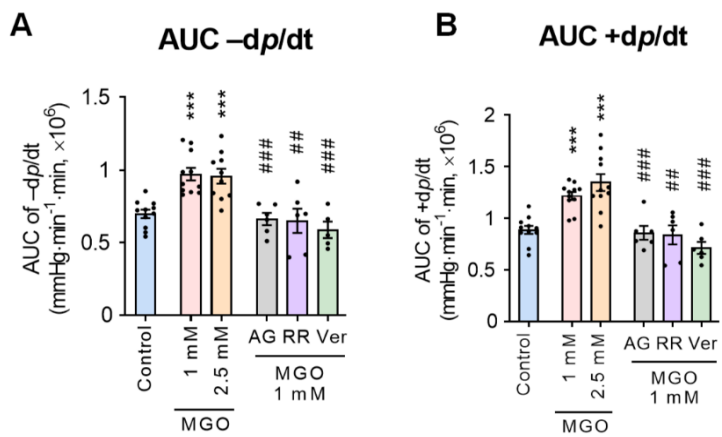

**Fig. S2.** Effect of MGO on cardiac haemodynamic response to stress test after 10 µM isoproterenol stimulation in isolated perfused hearts. Data shows the area under curve of positive differentials of left ventricular force development (+d*p*/dt) (A), and negative differentials of left ventricular force development (−d*p*/dt) (B) during the stabilisation period after the second stress test with 10 µM isoproterenol. After 20 min of stabilisation, hearts received either perfusate alone vehicle control; 1 mM MGO alone; 2.5 mM MGO alone; 1 mM MGO + 1 mM aminoguanidine (AG); 1 mM MGO + 2 µM ruthenium red (RR); or 1 mM MGO + 0.1 µM verapamil (Ver) for 95 min. Means ± SEM of 6-11 rats per group. *** $p < 0.001$: significantly different from control and ## $p < 0.01$; ### $p < 0.001$: significantly different from MGO 1 mM group; by Mann-Whitney comparison test.

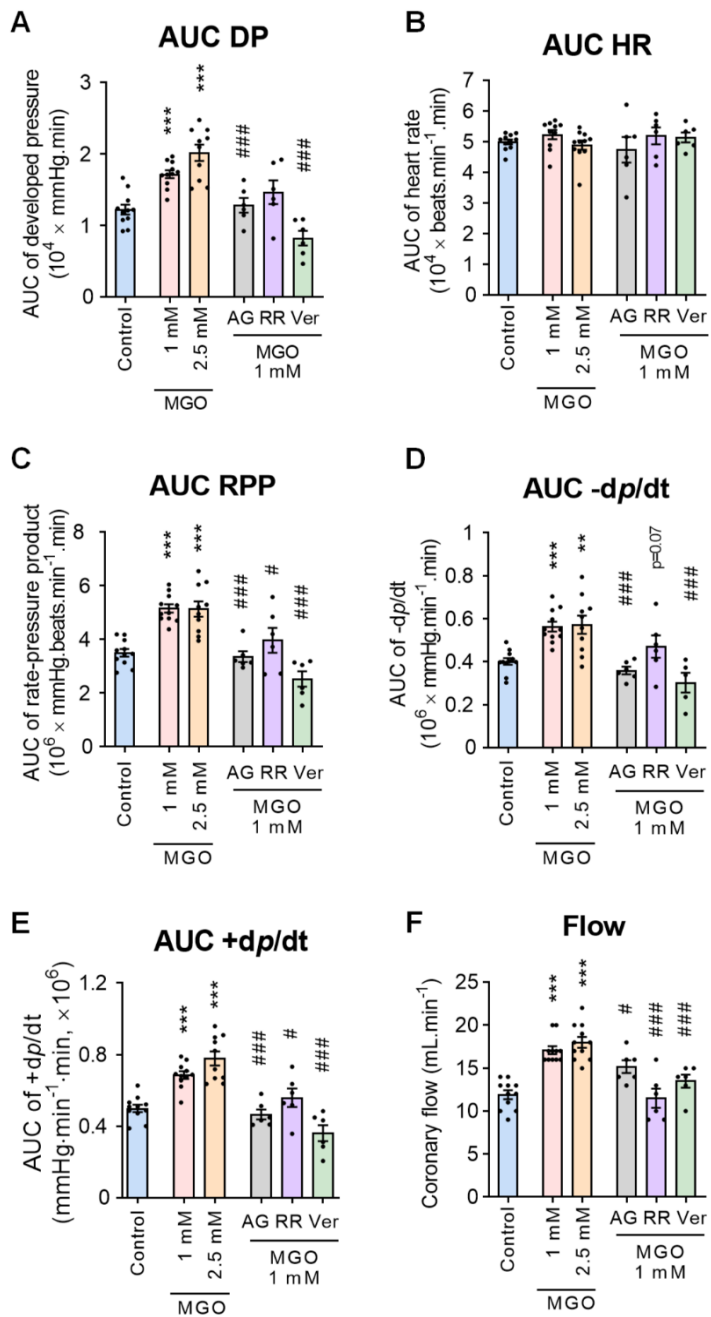

**Fig. S3.** Effect of MGO on cardiac haemodynamic response to stress test after 1 µM isoproterenol stimulation in isolated perfused hearts. Data shows the area under curve of each functional parameter, developed pressure (DP) (A), heart rate (HR) (B), rate-pressure product (RPP) (C), +d$p$/dt (D), –d$p$/dt (E), during the stabilisation period after the first stress test with 1 µM isoproterenol, and coronary flow (F) at the end of the time course (t115 min). After 20 min of stabilisation, hearts received either perfusate alone vehicle control; 1 mM MGO alone; 2.5 mM MGO alone; 1 mM MGO + 1 mM aminoguanidine (AG); 1 mM MGO + 2 µM ruthenium red (RR); or 1 mM MGO + 0.1 µM verapamil (Ver) for 95 min. Means ± SEM of 6-11 rats per group. ** $p < 0.01$; *** $p < 0.001$: significantly different from control and # $p < 0.05$; ## $p < 0.01$; ### $p < 0.001$: significantly different from MGO 1 mM group; by Mann-Whitney comparison test.

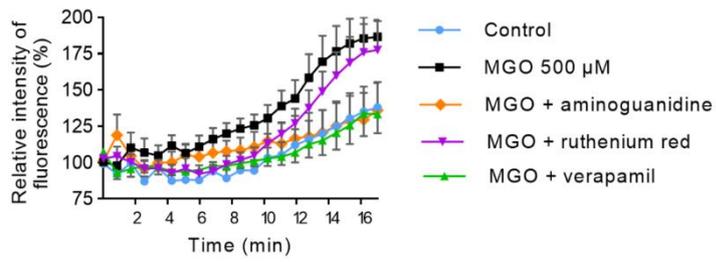

**Fig. S4.** Effect of aminoguanidine and calcium channels inhibitors on MGO-induced intracellular calcium levels increase. After 45 min-incubation with Fluo-4, fluorescence intensity was measured in fluorescence microscopy (ex 475 nm/ex 530 nm). After 1 min of acquisition, aminoguanidine at 500 µM, ruthenium red at 10 µM or verapamil at 1 µM was added to the medium, then at the third minute, MGO at 500 µM was added, and the acquisition was continued for 14 min.

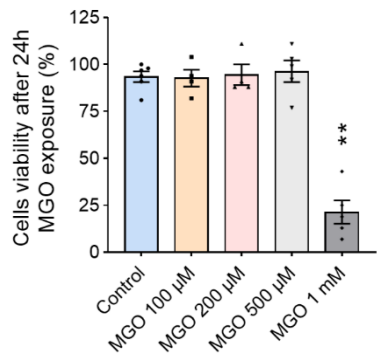

**Fig. S5.** Effect of MGO on cardiomyocytes viability. Neonatal rat cardiomyocytes were treated by MGO during 24 hours at 100 µM, 200 µM, 500 µM or 1 mM. Cells were stained by adding Prestoblue® followed by incubation for 30 min at 37 °C in an atmosphere of 95% $O_2$ / 5% $CO_2$. At the completion of sample incubation, fluorescence intensity was determined in a plate reader at a wavelength of 535 nm (excitation) and 590 nm (emission), then compared to fluorescence intensity determined before MGO treatment. Comparisons were made using Mann-Whitney test (means of 4-6 wells ± SEM; ** $p < 0.01$).

## 1. Cardiac function study in isolated perfused rat heart system

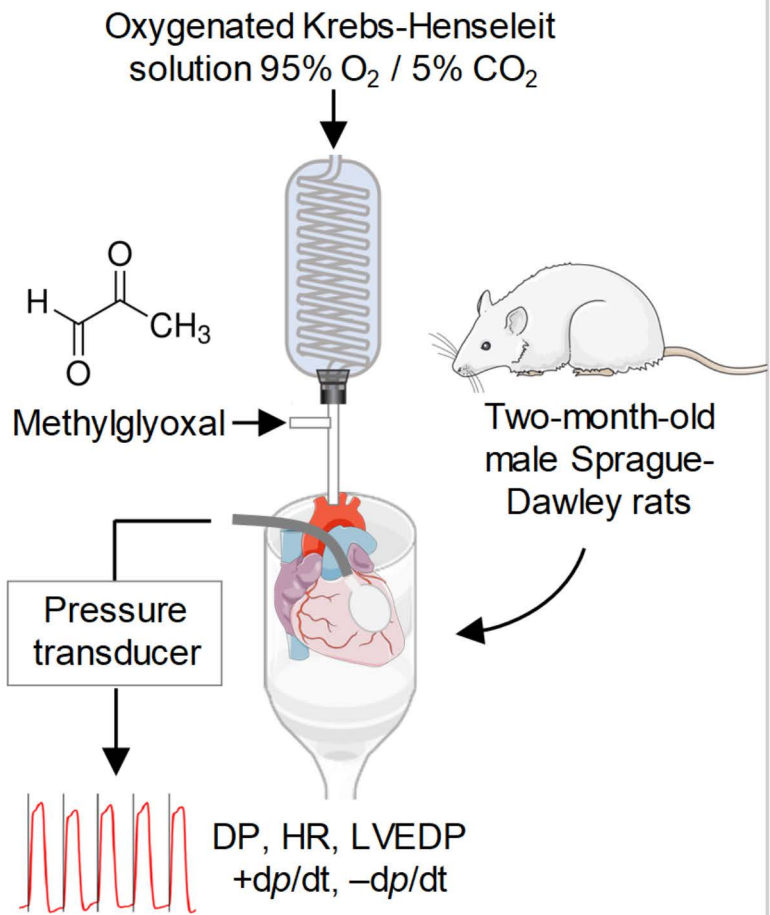

## 2. Real-time measurement of intracellular $Ca^{2+}$ using Fluo-4

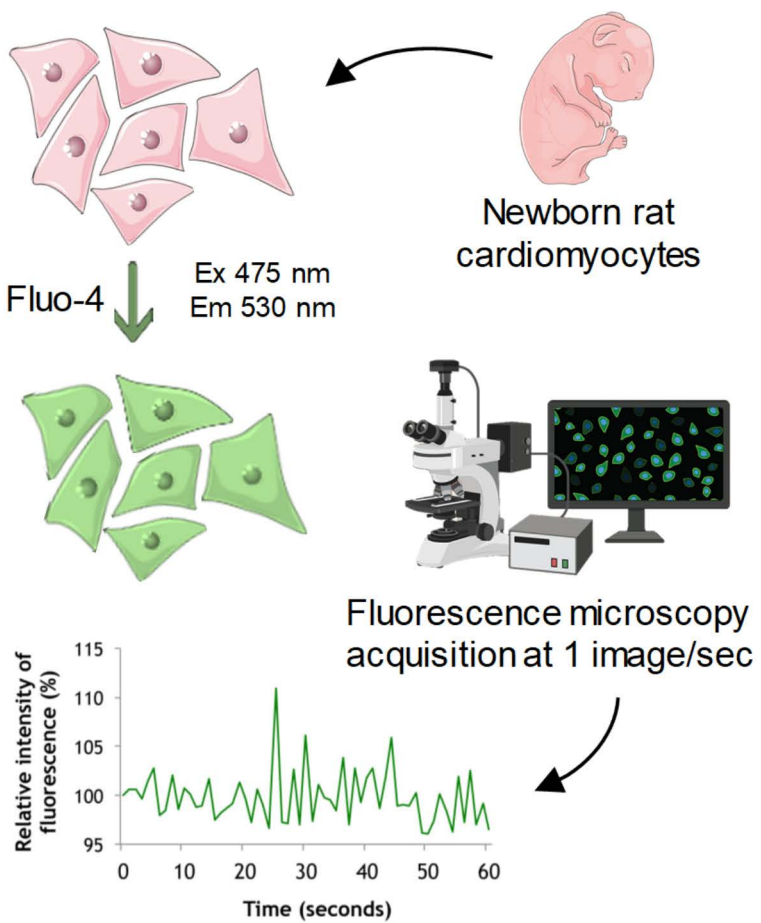

## 3. Conclusion

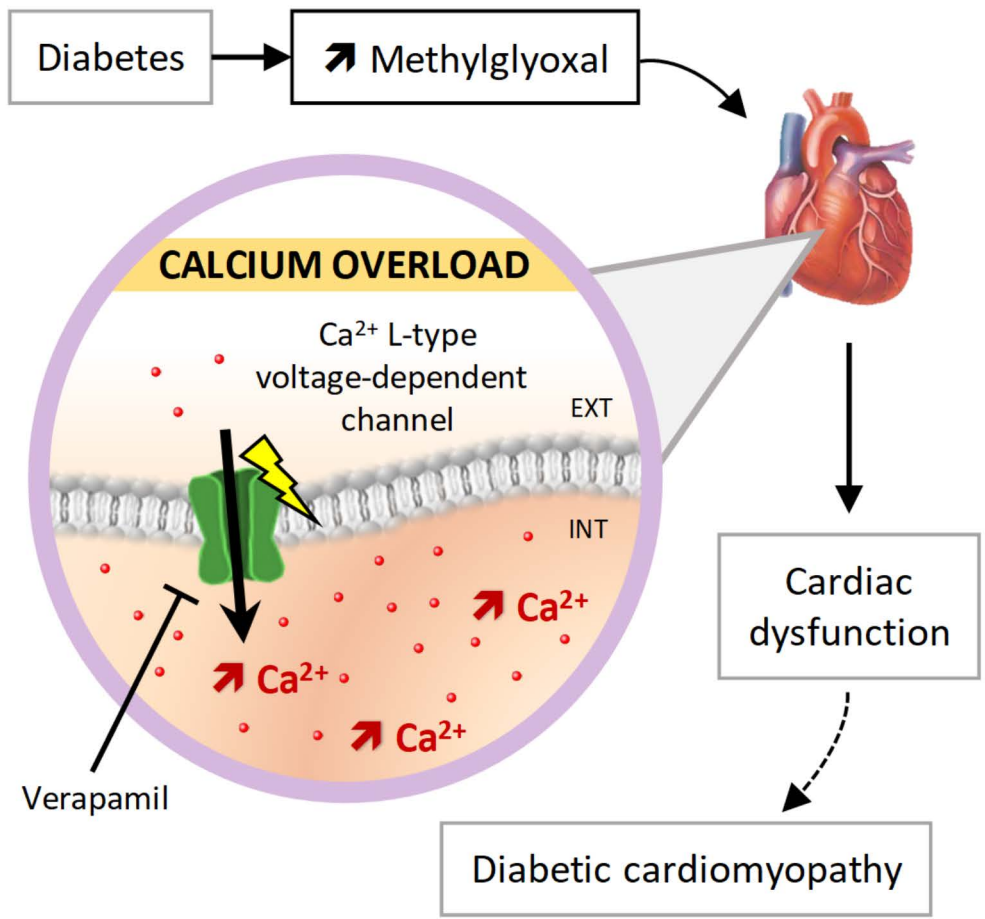

# HIGHLIGHTS

Methylglyoxal increases contractility in isolated perfused rat heart.

Methylglyoxal prolongs the positive inotropic effect of isoprenaline.

Methylglyoxal increases intracellular calcium levels in cardiomyocytes.

Methylglyoxal-induced cardiac dysfunction involves impairment of calcium channels.